\documentclass[10pt,aps,pre,notitlepage,onecolumn,superscriptaddress,preprintnumbers,amsmath,amssymb]{revtex4-1}

\usepackage{graphicx}
\usepackage{dcolumn}
\usepackage{bm}
\usepackage{color}

\begin{document}

\title{Evolutionary network games:\\ Equilibria from imitation and best-response dynamics} 

\author{Giulio Cimini}
\email{giulio.cimini@imtlucca.it}
\affiliation{IMT School for Advanced Studies, 55100 Lucca (Italy)}
\affiliation{Istituto dei Sistemi Complessi (ISC-CNR), 00185 Rome (Italy)}
\affiliation{Universidad Carlos III, 28911 Legan\'es, Madrid (Spain)}
\date{\today}

\begin{abstract}
We consider games of strategic substitutes and strategic complements on networks, and introduce two different evolutionary dynamics in order to refine their multiplicity of equilibria. 
We analyze the system through a mean field approach to both dynamics. We find that for the best-shot game, taken as  a representative example of strategic substitutes, a replicator-like (imitative) dynamics 
does not lead to Nash equilibria, whereas, it leads to a unique equilibrium (full cooperation or full defection, depending on the initial condition and the game parameters) for complements, 
represented by a coordination game. On the other hand, when the dynamics becomes more cognitively demanding in the form of a best response, predictions are always Nash equilibria 
(at least when individuals are fully rational): For the best-shot game we find a reduced set of equilibria with a definite value of the fraction of contributors, whereas, for the coordination game 
symmetric equilibria arise only for low or high initial fractions of cooperators. We further extend our analytical study by considering complex topologies through a heterogeneous mean field technique, 
and show that the nature of the selected equilibria does not change for the best-shot game. However, for coordination games we reveal an important difference, namely that on infinitely large scale-free networks 
cooperative equilibria arise for any value of the incentive to cooperate. Our analytical results are confirmed by numerical simulations available in the literature, 
and open the question of whether there can be a dynamics that consistently leads to stringent equilibria refinements for both classes of games.
\vspace{0.5cm}\\
{\em List of abbreviations used}: PI (proportional imitation), BR (best response), MF (homogeneous mean field), HMF (heterogeneous mean field)
\vspace{0.25cm}
\vspace{0.25cm}
\end{abstract}

\maketitle

\section{Introduction}\label{sec:intro}

Strategic interactions among individuals located on a network, be it geographical, social or of any other nature, are becoming increasingly relevant in many economic contexts. 
Decisions made by our neighbors on the network influence ours, and are in turn influenced by their other neighbors to whom we may or may not be connected. 
Such a framework makes finding the best strategy a very complex problem, almost always plagued by a very large multiplicity of equilibria. 
Researchers are devoting much effort to this problem, and an increasing body of knowledge is being consolidated \cite{goyal:2007,vega-redondo:2007,jackson:2008}. 
In this work we consider games of strategic substitutes and strategic complements on networks, as discussed in \cite{galeotti:2010}. 
In this paper, Galeotti {\em et al.}\ obtained an important reduction in the number of game equilibria by going from a complete information setting to an incomplete one. 
They introduced incomplete information by assuming that each player is only aware of the number of neighbors she has, but not of their identity nor of the number of neighbors they have in turn.
We here aim at providing an alternative equilibrium refinement by looking at network games from a evolutionary viewpoint. 
In particular, we look for the set of equilibria which can be accessed according to two different dynamics for players' strategies, and discuss the implications of such reduction. 
Furthermore, we go beyond the state of the art mean field approach and consider the role of complex topologies with an heterogeneous mean field technique. 

Our work belongs to the literature on strategic interactions in networks and its applications to economics 
\cite{angeletos:2007,ballester:2006,ballester:2010,bergemann:2009,bramouille:2011,calvo:2009,glaeser:2003,goyal:2001,vives:1999}. 
In particular, one of the games we study is a discrete version of a public goods game proposed by Bramoull\'e and Kranton \cite{bramouille:2007}, 
who opened the way to the problem of equilibrium selection in this kind of games under complete information. 
Bramoull\'e further considered this problem \cite{bram:2007} for the case of anti-coordination games on networks, showing that network effects are much stronger than for coordination games. 
As already stated, our paper originates from Galeotti {\em et al.} \cite{galeotti:2010}, for they considered one-shot games with strategic complements and substitutes 
and model equilibria resulting from incomplete information. Our approach is instead based on evolutionary selection of equilibria---pertaining to the large body of work 
emanating from the Nash programme \cite{Nash1950,MasColell1997,Fudenberg1998,young:1998}---and is thus complementary to theirs. 
In particular we focus on the analysis of two evolutionary dynamics (see Roca {\em et al.}\ \cite{roca:2009} for a review of the literature) 
in two representative games, and on how this dynamics leads to a refinement of the Nash equilibria or to other final states. 
The dynamics we consider are Proportional Imitation \cite{Helbing1992,schlag:1998}, that does not lead in general to Nash equilibria, 
and Best Response \cite{Matsui1992,Blume1993}, that instead allows for convergence to Nash equilibria---an issue about which there are a number of interesting results 
in the case of a well-mixed population \cite{Hart2000,Hart2003,Hart2011}. 
As we are working on a network setup, our specific perspective is close to that of Boncinelli and Pin \cite{boncinelli:2012}. 
They elaborate on the literature on stochastic stability \cite{young:1998,blume:2003} (see \cite{Blume1993,blume:1995} for an early example of related dynamics on lattices) 
as a device that selects the equilibria that are more likely to be observed in the long run, in the presence of small errors occurring with a vanishing probability. 
They work from the observation \cite{bergin:1996} that different equilibria can be selected depending on assumptions on the relative likelihood of different types of errors. 
Thus, Boncinelli and Pin work with a Best Response dynamics and by means of a Markov Chain analysis find, counter-intuitively, that when contributors are the most perturbed players, 
the selected equilibrium is the one with the highest contribution. The techniques we use here are based on differential equations and have a more dynamical character, and we do not incorporate 
the possibility of having special distributions of errors---although we do consider random mistakes. 
Particularly relevant to our work is the paper by L\'opez-Pintado \cite{dunia:2007} (see \cite{dunia:2013} for an extension to the case of directed networks) where a mean field dynamical approach 
involving a random subsample of players is proposed. Within this framework, the network is dynamic, as if at each period the network were generated randomly. 
Then a unique globally stable state of the dynamics is found, although the identities of free riders might change from one period to another. 
The difference with our work is that we do not deal with a time-dependent subsample of the population, but we use a global mean field approach (possibly depending on the connectivity of individuals) 
to describe the behavior of a static network.

In the remainder of this Introduction we present the games we study and the dynamics we apply for equilibrium refinement in detail, discuss the implications of such a framework 
on the informational settings we are considering, and summarize our main contributions.

\subsection{Framework}

\subsubsection{Games}

We consider a finite set of agents $I$ of cardinality $n$, linked together in a fixed, undirected, exogenous network. 
The links between agents reflect social interactions, and connected agents are said to be ``neighbors''. The network is defined through a $n \times n$ symmetric matrix $G$ with null diagonal, 
where $G_{i j} = 1$ means that agents $i$ and $j$ are neighbors, while $G_{i j} = 0$ that they are not. We indicate with $N_i$ the set of $i$'s neighbors,
\emph{i.e.}, $N_i = \{j \in I: G_{i j} = 1\}$, where the number of such neighbors $|N_i|=k_i$ is the \emph{degree} of the node. 

Each player can take one of two actions $X=\{0,1\}$, with $x_i\in X$ denoting $i$'s action. Hence, only pure strategies are considered. 
In our context (particularly for the case of substitutes), action 1 may be interpreted as \emph{cooperating} and action 0 as not doing so---or \emph{defecting}. 
Thus, the two actions are labeled in the rest of the paper as $C$ and $D$, respectively. 
There is a cost $c$, where $0<c<1$, for choosing action $x=1$, while action $x=0$ bears no cost.

In what follows we concentrate on two games, the best-shot game and a coordination game, as representative instances of strategic substitutes and strategic complements, respectively. 
We choose specific examples for the sake of being able to study analytically their dynamics. 
To define the payoffs we introduce the following notation: $x_{N_i}=\sum_{j\in N_i}x_j$ is the aggregate action in $N_i$ and $y_i=x_i+x_{N_i}$. 

\paragraph{Strategic Substitutes: Best-shot game}
This game was first considered by Bramoull\'e and Kranton \cite{bramouille:2007} as a model of the local provision of a public good. 
As stated above, we consider the discrete version, where there are only two actions available, as in \cite{galeotti:2010,boncinelli:2012}. The corresponding payoff function takes the form
\begin{equation}\label{eq.SS}
\pi_i=\Theta_H(y_i-1)-c\,x_i
\end{equation}
where $\Theta_H(\cdot)$ is the Heaviside step function $\Theta_H(x)=1$ if $x\ge0$ and $\Theta_H(x)=0$ otherwise.

\paragraph{Strategic Complements: Coordination game}

For our second example, we follow Galeotti {\em et al.} \cite{galeotti:2010} and consider again a 
discrete version of the game, but now let the payoffs of any particular agent $i$ be given by
\begin{equation}\pi_i=(\alpha x_{N_i}-c)\,x_i. \label{eq:SC}
\end{equation} Assuming that $c > \alpha > 0$, we are faced with a coordination
game where, as discussed in \cite{galeotti:2010}, depending on the underlying network and the information conditions, there can generally be multiple equilibria.

\subsubsection{Dynamics}

Within the two games we have presented above, we now consider evolutionary dynamics for players' strategies. 
Starting at $t=0$ with a certain fraction $\rho(0)=\sum_ix_i(0)/n$ of players randomly chosen to undertake action $x=1$, at each round $t$ of the game 
players collect their payoff $\pi(t)$ according to their neighbors' actions and the kind of game under consideration. Subsequently, a fraction $q$ of players update their strategy. 
We consider two different mechanisms for strategy updating:

\paragraph{Proportional Imitation} (PI) \cite{Helbing1992,schlag:1998}. It represents a rule of imitative nature in which player $i$ may copy the strategy of a selected counterpart $j$, 
which is chosen randomly among the $k_i$ neighbors of $i$. The probability that $i$ copies $j$'s strategy depends on the difference between the payoffs they obtained in the previous round of the game:
\begin{equation}\label{eq.PI}
\mathcal P\left\{x_j(t)\rightarrow x_i(t+1)\right\}=\begin{cases}
	[\pi_j(t)-\pi_i(t)]/\Phi&\mbox{if $\pi_j(t)>\pi_i(t)$}\\
	\epsilon&\mbox{otherwise}
	\end{cases}
\end{equation}
where $\Phi$ is a normalization constant that ensures $\mathcal P\{\cdot\}\in[0,1]$, and $0\le\epsilon<1$ allows for mistakes (\emph{i.e.}, copying an action that yielded less payoff in the previous round). 
Note that because of the imitation mechanism of PI,  the configurations $x_i=1$ $\forall i$ and $x_i=0$ $\forall i$ are absorbing states: 
The system cannot escape from them and not even mistakes can re-introduce strategies, as they always involve imitation. On the other hand, it can be shown that PI is equivalent 
to the well-known replicator dynamics in the limit of an infinitely large, well-mixed population (equivalently, on a complete graph) \cite{Gintis2009,Levine2007}. 
As was first put by Schlag \cite{schlag:1998}, the assumption that agents play a random-matching game  in a large population and learn the actual payoff of another randomly chosen agent, 
along with a rule of action that increases their expected payoff, leads to a probability of switching to the other agent's strategy that is proportional to the difference in payoffs. 
The corresponding aggregate dynamics is like the replicator dynamics. See also \cite{borgers:1997} for another interpretation of these dynamics in terms of learning. 

\paragraph{Best Response} (BR). This rule was introduced in \cite{Matsui1992,Blume1993} and has been widely used in the economics literature. 
BR describes players that are rational and choose their strategy (myopically) in order to maximize their payoff, assuming that their neighbors will again do what they did in the last round.  
This means that each player $i$, given the past actions of their partners $x_{N_i}(t)$, computes the payoffs that she would obtain by choosing action 1 (cooperating) or 0 (defecting) at time $t$, 
respectively $\tilde{\pi}_C(t)$ and $\tilde{\pi}_D(t)$.  Then actions are updated as follows:
\begin{equation}\label{eq.BR}
\mathcal P\left\{x_i(t+1)=1\right\}=\begin{cases}
	1-\epsilon&\mbox{if $\tilde{\pi}_C(t)>\tilde{\pi}_D(t)$}\\
	\epsilon&\mbox{if $\tilde{\pi}_C(t)<\tilde{\pi}_D(t)$}
	\end{cases}
\qquad;\qquad
\mathcal P\left\{x_i(t+1)=0\right\}=\begin{cases}
	\epsilon&\mbox{if $\tilde{\pi}_C(t)>\tilde{\pi}_D(t)$}\\
	1-\epsilon&\mbox{if $\tilde{\pi}_C(t)<\tilde{\pi}_D(t)$}
	\end{cases}
\end{equation}
and $x_i(t+1)=x_i(t)$ if $\tilde{\pi}_C(t)=\tilde{\pi}_D(t)$. Here again $0\le\epsilon<1$ represents the probability of making a mistake, with $\epsilon=0$ indicating fully rational players. 

\medskip

The reason to study these two dynamics is because they may lead to different results as they represent very different evolutions of the players' strategies. 
In this respect, it is important to mention that, in the case $\epsilon=0$, Nash equilibria are stable by definition under BR dynamics and, vice-versa, any stationary state found by BR 
is necessarily a Nash equilibrium. On the contrary, with PI this is not always true: Even in the absence of mistakes, players can change action by copying better-performing neighbors, 
also if such change leads to a decreasing of their payoffs in the next round. Another difference between the two dynamics is the amount of cognitive capability they assume for the players: 
Whereas PI refers to agents with very limited rationality, that imitate a randomly chosen neighbor on the only condition that she does better, BR requires agents with a much more developed analytic ability.

\subsubsection{Analytical and informational settings}

We study how the system evolves by either of these two dynamics, starting from an initial random distribution of strategies. 
In particular, we are interested in the global fraction of cooperators $\rho(t)=\sum_ix_i(t)/n$ and its possible stationary value $\rho_s$.
We carry out our calculations in the framework of a homogeneous mean field (MF) approximation, 
which is most appropriate to study networks with homogeneous degree distribution $P(k)$ like Erd\"{o}s-R\'{e}nyi random graphs \cite{Erdos1960}. 
The basic assumption underlying this approach is that every player interacts with an ``average player'' that represents the actions of her neighbors. 
More formally, the MF approximation consists in assuming that when a player interacts with a neighbor of hers, 
the action of such a neighbor is $x=1$ with probability $\rho$ (and $x=0$ otherwise), independently on the particular pair of players considered \cite{MFgames}.
Loosely speaking, this amounts to having a very incomplete information setting, in which all players know only how many other players they will engage with, 
and is reminiscent of that used by Galeotti {\em et al.} \cite{galeotti:2010} for their refinement of equilibria. However, the analogy is not perfect and therefore, for the sake of accuracy, 
we do not dwell any further on the matter. In any case, MF represents our setup for most of the paper. 

As an extension of the results obtained in the above context, we also study the case of highly heterogeneous networks, \emph{i.e,}, networks with broad degree distribution $P(k)$, 
such as scale-free ones \cite{Barabasi1999}. In these cases in fact there are a number of players with many neighbors (``hubs'') and many players with only a few neighbors, 
and this heterogeneity may give rise to very different behaviors as compared to Erd\"{o}s-R\'{e}nyi systems. Analytically, this can be done by means of the 
heterogeneous mean field technique (HMF)~\cite{PastorSatorras2001} which generalizes, for the case of networks with arbitrary degree distribution, 
the equations describing the dynamical process by considering degree-block variables grouping nodes within the same degree. 
More formally, now when a player interacts with a neighbor of hers, the action of such a neighbor is $x=1$ with probability $\rho_k$ (and $x=0$ otherwise) 
if $k$ is the neighbor's degree ($\rho_k$ is the density of cooperators within players of degree $k$).
By resorting to this second perspective we are able to gain insights on the effects of heterogeneity on the evolutionary dynamics of our games. 

\subsection{Our contribution}

Within this framework, our main contribution can be summarized as follows. In our basic setup of homogeneous networks (described by the mean field approximation): 
For the best-shot game, PI leads to a stationary state in which all players play $x_i=0$, \emph{i.e.}, to full defection, which is 
however non-Nash as any player surrounded by defectors would obtain higher payoff by choosing cooperation (at odds with the standard version of the public goods game).
This is the result also in the presence of mistakes, unless the probability of errors becomes large, in which case the stationary state is the opposite, $x_i=1$, \emph{i.e.}, full cooperation, also non-Nash. 
Hence, PI does not lead to any refinement of the Nash equilibrium structure. On the contrary, BR leads to Nash equilibria characterized by a finite fraction of cooperators $\rho_s$, whereas, 
in the case when players are affected by errors, this fraction coincides with the probability of making an error as the mean degree of the network goes to infinity. 
The picture is different for the coordination game. In this case, PI does lead to Nash equilibria, selecting the coordination in 0 below a threshold value of $\alpha$ and the opposite state otherwise. 
This threshold is found to depend on the initial fraction $\rho(0)$ of players choosing $x=1$. Mistakes lead to the appearance of a new possibility, an intermediate value of the fraction of players choosing 1, 
and as before the initial value of this fraction governs which equilibrium is selected. BR gives similar results, albeit for the fact that a finite fraction of 1 actions can also be found even without mistakes, 
and with mistakes the equilibria are not full 0 or 1 but there is always a fraction of mistaken players. Finally, changing the analytical setting by proceeding to the heterogeneous mean field approach 
does not lead to any dramatic change in the structure of the equilibria for the best-shot game. Interestingly, things change significantly 
for coordination games---when played on infinitely large scale-free networks. In this case, which is the one where the heterogeneous mean field should make a difference, 
equilibria with non-vanishing cooperation obtain for any value of the incentive to cooperate (represented by the parameter $\alpha$).

The paper is organized in seven sections including this introduction. Section \ref{sec:bs_g} presents our analysis and results for the best-shot game. Section \ref{sec:c_g} deals with the coordination game. 
In both cases, the analytical framework is that of the mean field technique. After an overall analysis of global welfare performed in Section \ref{sec:welfare}, Section \ref{sec:hmf} presents the extensions 
of the results for both games within the heterogeneous mean field approach, including some background on the formalism itself. 
Finally, Section \ref{sec:simu} contains an assessment of the validity of all these analytical findings in light of the results of recent numerical simulations of the system described above, 
and Section \ref{sec:end} concludes summarizing our most important findings concerning the refinement of equilibria in network games and pointing to relevant open questions.

\section{Best-shot game}\label{sec:bs_g}

\subsection{Proportional imitation}

We begin by considering the case of strategic substitutes when imitation of a neighbor is only possible if she has obtained better payoff than the focal player, 
\emph{i.e.}, $\epsilon=0$ in eq.\ (\ref{eq.PI}). In that case, the main result is the following:

\medskip

{\bf Proposition 1.} 
Within the mean field formalism, under PI dynamics, when $\epsilon=0$ the final state for the population is the absorbing state with a density of cooperators $\rho=0$ (full defection) 
except if the initial state is full cooperation. 

\medskip

{\em Proof.} 
{Working in a mean field context means that individuals are well-mixed, \emph{i.e.}, every player interacts with average players. 
In this case the differential equation for the density of cooperators $\rho$ is}
\begin{equation}\label{eq.SS_PI}
\dot{\rho}/q=(1-\rho)\rho\,\mathcal{P}_{D\rightarrow C}-\rho(1-\rho)\,\mathcal{P}_{C\rightarrow D}.
\end{equation}
{The first term is the probability $(1-\rho)\rho$ of picking a defector with a neighboring cooperator, times the probability of imitation $\mathcal{P}_{D\rightarrow C}$. 
The second term is the probability $\rho(1-\rho)$ of picking a cooperator with a neighboring defector, times the probability of imitation $\mathcal{P}_{C\rightarrow D}$. 
In the best-shot game a defector cannot copy a neighboring cooperator (who has lower payoff by construction), whereas, a cooperator eventually copies one of her neighboring defectors 
(who has higher payoff). Hence $\mathcal{P}_{D\rightarrow C}=0$ and $\mathcal{P}_{C\rightarrow D}$ is equal to the payoff difference $1-(1-c)=c$. Since the normalization constant $\Phi=1$ 
for strategic substitutes, eq.\ (\ref{eq.SS_PI}) becomes}
\begin{equation}\label{eq.SS_PI_bis}
\dot{\rho}/q=-c\rho(1-\rho)
\end{equation}
The solution, for any initial condition $0<\rho(0)<1$, is
\begin{equation}\label{eq.SS_PI_sol}
\rho(t)=\{1+[\rho(0)^{-1}-1]e^{cqt}\}^{-1},
\end{equation}
hence $\rho(t)\rightarrow0$ for $t\rightarrow\infty$: The only stationary state is full defection unless $\rho(0)=1$. $\blacksquare$

\medskip

{\bf Remark 1:} As discussed above, PI does not necessarily lead to Nash equilibria as asymptotic, stationary states. This is clear in this case. 
For any $\rho(0)<1$ the population ends up in full defection, even if every individual player would be better off by switching to cooperation. 
This phenomenon is likely to arise from the substitutes or anti-coordination character of the game: In a context in which it is best to do the opposite of the other players, 
imitation does not seem the best way for players to decide on their actions.

\bigskip

{\bf Proposition 2.} 
Within the mean field formalism, under PI dynamics, when $\epsilon\in(0,1)$ the final state for the population is the absorbing state  $\rho=0$ (full defection) 
when $\epsilon < c$, $\rho=\rho(0)$ when $\epsilon=c$, and $\rho=1$ when $\epsilon>c$. When the initial state is $\rho(0)=0$ or $\rho(0)=1$, it remains unchanged.  

\medskip

{\em Proof.} 
Eq.\ (\ref{eq.SS_PI}) is still valid, with $\mathcal{P}_{C\rightarrow D}$ unchanged, whereas, $\mathcal{P}_{D\rightarrow C}=\epsilon$. 
By introducing the effective cost $\tilde{c}=c-\epsilon$ we can rewrite eq.\ (\ref{eq.SS_PI_sol}) as:
\begin{equation}\label{eq.SS_PI_sol_e}
\rho(t)=\{1+[\rho(0)^{-1}-1]e^{\tilde{c}qt}\}^{-1}
\end{equation}
Hence $\rho(t)\rightarrow0$ for $t\rightarrow\infty$ only for $\tilde{c}>0$ ($\epsilon<c$); 
Instead for $\tilde{c}=0$ ($\epsilon=c$) then $\rho(t)\equiv\rho(0)$ $\forall t$, and for $\tilde{c}<0$ ($\epsilon>c$) then $\rho(t)\rightarrow1$ for $t\rightarrow\infty$ (cooperation is favored now).
$\blacksquare$

\medskip

{\bf Remark 2:} As before, PI does not  drive the population to a Nash equilibrium, independently of the probability of making a mistake. However, mistakes do introduce a bias towards cooperation 
and thus a new scenario: When their probability exceeds the cost of cooperating, the whole population ends up cooperating. 

\subsection{Best response}

We now turn to the case of the best response dynamics, which (at least for $\epsilon=0$) is guaranteed to drive the system towards Nash equilibria. 
In this scenario, we have not been able to find a rigorous proof of our main result, but we can make some approximations in the equation that support it. 
As we will see, our main conclusion is that, within the mean field formalism under BR dynamics, when $\epsilon=0$ the final state for the population is a mixed state $\rho=\rho_s$, $0<\rho_s<1$, 
for any initial condition. 

\medskip

Indeed, for BR dynamics without mistakes,  
the homogeneous mean field equation for $\dot{\rho}$ is
\begin{equation}\label{eq.SS_BR}
\dot{\rho}/q=-\rho\,Q[\pi_C<\pi_D]+(1-\rho)\,Q[\pi_C>\pi_D]
\end{equation}
where the first term is the probability of picking a cooperator who would do better by defecting, 
and the second term is the probability of picking a defector who would do better by cooperating. This far, no approximation has been made; 
However, these two probabilities cannot be exactly computed and we need to estimate them. 

To evaluate the two probabilities, we can recall that $\pi_C=1-c$ always, whereas, $\pi_D=0$ when none of the neighbors cooperates and $\pi_D=1$ otherwise. 
Therefore, for an average player of degree $k$ we have that $Q_k[\pi_C>\pi_D]=(1-\rho)^k$. Consistently with the mean field framework we are working on, 
as a rough approximation we can assume that every player has degree $\bar{k}$ (the average degree of the network), so that $Q[\pi_C>\pi_D]=1-Q[\pi_C<\pi_D]=(1-\rho)^{\bar{k}}$. Thus, we have
\begin{equation}\label{eq.SS_BR_1}
\dot{\rho}/q=(1-\rho)^{\bar{k}}-\rho.
\end{equation}
To go beyond this simple estimation, we can work out a better approximation by integrating $Q_k[\pi_C>\pi_D]$ over the probability distribution of players' degrees $P(k)$. 
For Erd\"{o}s-R\'{e}nyi random graphs, in the limit of large populations ($n\rightarrow\infty$), {it is $P(k)\simeq\bar{k}^k e^{-\bar{k}}/k!$}. This leads to $Q[\pi_C>\pi_D]=e^{-\bar{k}\rho}$ and, subsequently,
\begin{equation}\label{eq.SS_BR_2}
\dot{\rho}/q=e^{-\bar{k}\rho}-\rho.
\end{equation}

\medskip

{\bf Remark 3.} The precise asymptotic value for the density of cooperators, $\rho_s$, depends on the approximation considered above. 
However, at least for networks that are not too inhomogeneous, the approximations turn out to be very good, 
and therefore the corresponding picture for the evolution of the population is quite accurate. 
It is interesting to note that, whatever its exact value, in both cases  $\rho_s$ is such that the right-hand sides of eq.\ (\ref{eq.SS_BR_1}) and eq.\ (\ref{eq.SS_BR_2}) 
vanishes and, furthermore, $\rho_s$ is an attractor of the dynamics, because $d(\dot{\rho})/d\rho|_{\rho_s}<0$. 

\bigskip

How is the above result modified by mistakes? When $\epsilon\in(0,1)$, eq.\ (\ref{eq.SS_BR}) becomes:
\begin{eqnarray}\label{eq.SS_BR_e}
\dot{\rho}/q&=&Q[\pi_C<\pi_D]\{-\rho(1-\epsilon)+(1-\rho)\epsilon\}+Q[\pi_C>\pi_D]\{(1-\rho)(1-\epsilon)-\rho\epsilon\}\nonumber\\
&=&Q[\pi_C<\pi_D](-\rho+\epsilon)+Q[\pi_C>\pi_D](1-\rho-\epsilon)
\end{eqnarray}
where the first term accounts for cooperators rightfully switching to defection and defectors wrongly selecting cooperation, 
while the second term accounts for defectors correctly choosing cooperation and cooperators mistaken to defection. 
Proceeding as before, and again in the limit $n\rightarrow\infty$, we approximate $Q[\pi_C>\pi_D]=e^{-\bar{k}\rho}$,  thus arriving at
\begin{equation}\label{eq.SS_BR_2_e}
\dot{\rho}/q=(1-2\epsilon)e^{-\bar{k}\rho}-(\rho-\epsilon)
\end{equation}
from which it is possible to find the attractor of the dynamics $\rho_s$. Such attractor in turn exists if $$\epsilon<(1+\bar{k}^{-1}e^{\bar{k}\rho_s})/2,$$
a threshold that is bounded below by 1/2, which would be tantamount to players choosing their action at random. Therefore, all reasonable values for the probability of errors allow for equilibria.  

\begin{figure}[t!]
\centering
\includegraphics[width=0.5\textwidth]{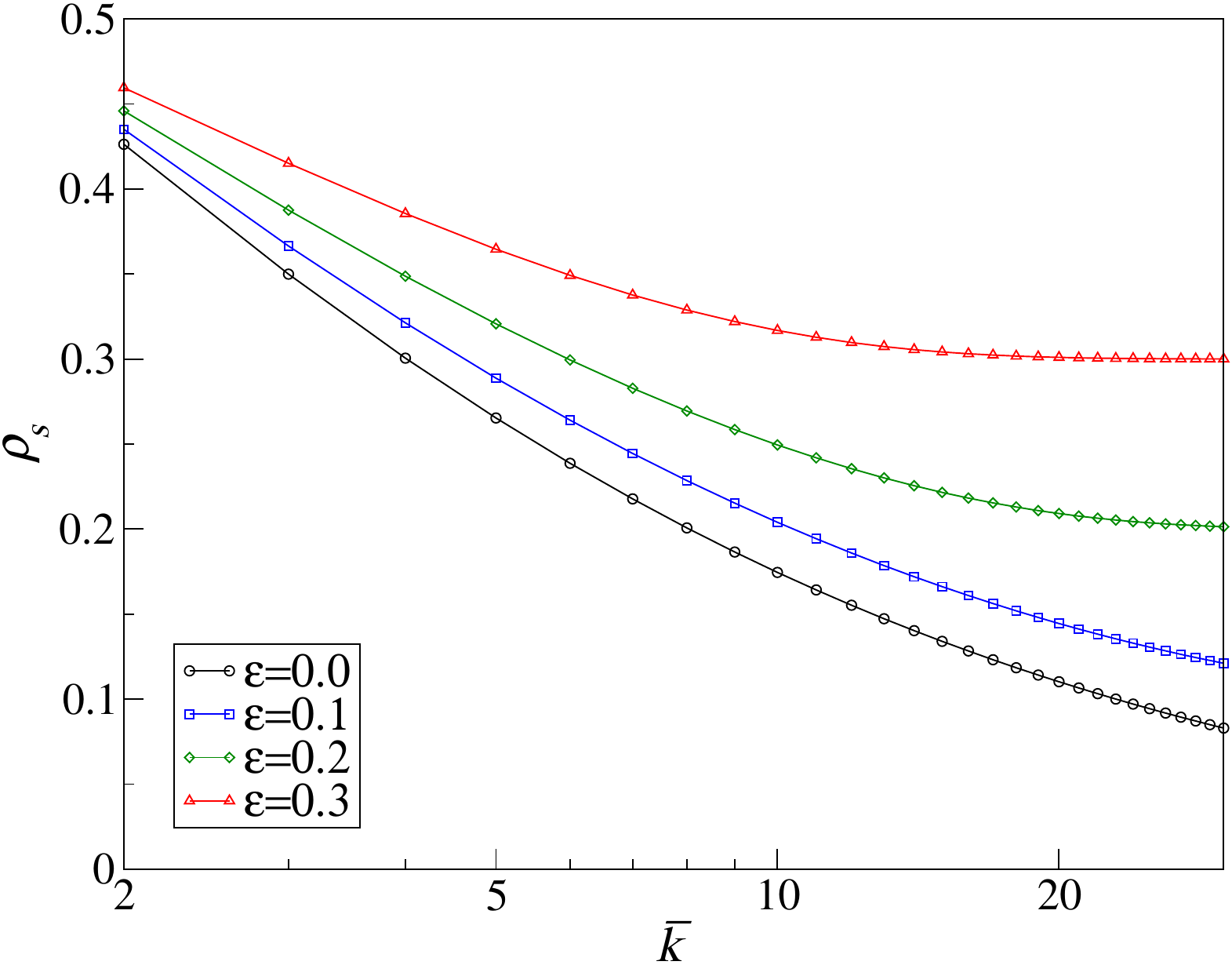}
\caption{Best-shot game under BR dynamics in the mean field framework. Shown are the asymptotic cooperation values $\rho_s$ vs the average degree $\bar{k}$ for different values 
of the probability of making a mistake $\epsilon$. Values are obtained by numerically solving eq.\ (\ref{eq.SS_BR_2_e}).}\label{fig.SS_BR_rc}
\end{figure}

\medskip

{\bf Remark 4.}
To gain some insight on the cooperation levels arising from BR dynamics in the Nash equilibria, we have numerically solved eq.\ (\ref{eq.SS_BR_2_e}). The values for $\rho_s$ are plotted in 
Fig.\ \ref{fig.SS_BR_rc} for different values of $\epsilon$, as a function of $\bar{k}$. We observe that the larger $\bar{k}$, the lower the cooperation level. 
The intuition behind such result is that the more connections every player has, the lower the need to play 1 to ensure obtaining a positive payoff. 
It could then be thought that this conclusion is reminiscent of the equilibria found for best-shot games in \cite{galeotti:2010}, which are non increasing in the degree. 
However, this is not the case, as in our work we are considering an iterated game that can perfectly lead to high degree nodes having to cooperate. 
Note also that this approach leads to a definite value for the density of cooperators in the Nash equilibrium, but there can be many action profiles for the player compatible with that value, 
so multiplicity of equilibria is reduced but not suppressed. 

\medskip

{\bf Remark 5.} 
From Fig.\ \ref{fig.SS_BR_rc} it is also apparent that as the likelihood of mistakes increases, the density of cooperators at equilibrium increases. 
Note that for very large values of the connectivity $\bar{k}$, eq.\ (\ref{eq.SS_BR_2_e}) has solution $\rho(t)=\rho(0)e^{-qt}+\epsilon$, 
and thus $\rho_s\equiv\epsilon$, in agreement with the fact that when a player has many neighbors 
she can assume that a fraction $\epsilon$ of them will cooperate, thus turning defection into her BR.

\section{Coordination game}\label{sec:c_g}

We now turn to the case of strategic complements, exemplified by our coordination game. As above, we start from the case without mistakes, and we subsequently see how they affect the results.

\subsection{Proportional imitation}

\medskip

{\bf Proposition 3.} 
Within the mean field formalism, under PI dynamics, when $\epsilon=0$ the final state for the population is the absorbing state with a density of cooperators $\rho=0$ (full defection) 
when $\alpha<\alpha_c\equiv c/[\bar{k}\rho(0)]$, and the absorbing state with $\rho=1$ when $\alpha>\alpha_c$. In the case $\alpha=\alpha_c$ both outcomes are possible. 

\medskip

{\em Proof.} 
Still within our homogeneous mean field context, the differential equation for the density of cooperators $\rho$ is again eq.\ (\ref{eq.SS_PI}). 
As we are in the case in which $\epsilon=0$, we have that $\mathcal{P}_{D\rightarrow C}=(\pi_C-\pi_D)Q[\pi_C>\pi_D]/\Phi$ and 
$\mathcal{P}_{C\rightarrow D}=(\pi_D-\pi_C)Q[\pi_C<\pi_D]/\Phi=-(\pi_C-\pi_D)(1-Q[\pi_C>\pi_D])/\Phi$, 
where for strategic complements $\Phi=\alpha k_{max}$. Given that $\pi_D=0$ and that, consistently with our MF framework, $\pi_C=\alpha\bar{k}\rho-c$, we find:
\begin{equation}\label{eq.SC_PI_sol}
\dot{\rho}/q=\rho(1-\rho)(\alpha\bar{k}\rho-c)/\Phi=c\rho(1-\rho)(\rho/\rho_c-1)/\Phi,
\end{equation}
where we have introduced the values $\rho_c=\rho(0)[\alpha_c/\alpha]$ and $\alpha_c\equiv c/[\bar{k}\rho(0)]$. 

It is easy to see that $\rho=\rho_c$ is an unstable equilibrium, 
as $\dot{\rho}<0$ for $\rho<\rho_c$ and $\dot{\rho}>0$ for $\rho>\rho_c$. 
Therefore, we have two different cases: When $\alpha>\alpha_c$ then $\rho_c<\rho(0)$ and the final state is full cooperation ($\rho=1$), 
whereas, when $\alpha<\alpha_c$ then $\rho_c>\rho(0)$ and the outcome is full defection ($\rho=0$). 
When $\alpha\equiv\alpha_c$ then $\rho_c\equiv\rho(0)$, so both outcomes are in principle possible. $\blacksquare$

\medskip

{\bf Remark 6.}
The same (but opposite) intuition we discussed in Remark 1 about the outcome of PI on substitute games suggests that imitation is indeed a good procedure to choose actions in a coordination setup. 
In fact, contrary to the case of the best-shot game, in the coordination game PI does lead to Nash equilibria, and indeed it makes a very precise prediction: 
A unique equilibrium that depends on the initial density. Turning around the condition for the separatrix, we have $\rho(0)<c/(\bar{k}\alpha)$, \emph{i.e.}, 
when few people cooperate initially then evolution leads to everybody defecting, and vice versa. In any event, having a unique equilibrium (except exactly at the separatrix) is a remarkable achievement. 

\medskip

{\bf Remark 7.} 
In a system where players may have different degrees, while full defection is always a Nash equilibrium for the coordination game, full cooperation becomes a Nash equilibrium only when $\alpha>c/k_{min}$, 
where $k_{min}$ is the smallest degree in the network---which means that only networks with $k_{min}>c/\alpha>1$ feature a fully cooperative Nash equilibrium. 

\bigskip

When $\epsilon\in(0,1)$, the problem becomes much more involved and we have not been able to prove rigorously our main result.  
In fact, now we have $\mathcal{P}_{D\rightarrow C}=(\pi_C-\pi_D)Q[\pi_C>\pi_D]/\Phi+\epsilon Q[\pi_C<\pi_D]$ and 
$\mathcal{P}_{C\rightarrow D}=(\pi_D-\pi_C)Q[\pi_C<\pi_D]/\Phi+\epsilon Q[\pi_C>\pi_D]$. 
Eq.\ (\ref{eq.SC_PI_sol}) thus becomes
\begin{equation}\label{eq.SC_PI_sol_e}
\dot{\rho}/q=\rho(1-\rho)\{c(\rho/\rho_c-1)/\Phi+\epsilon(1-2Q[\rho>\rho_c])\}
\end{equation}
where we have used $Q[\pi_C>0]\simeq Q[\alpha\bar{k}\rho>c]=Q[\rho>\rho_c]$. We then have three different cases which we can treat approximately:
\begin{itemize}
\item
When $\rho\simeq\rho_c$ then $Q[\rho>\rho_c]\simeq1/2$ and eq.\ (\ref{eq.SC_PI_sol_e}) reduces to eq.\ (\ref{eq.SC_PI_sol}), \emph{i.e.}, we would recover the result for the case with no mistakes. 
\item
When $\rho\gg\rho_c$ then $Q[\rho>\rho_c]\simeq1$ and eq.\ (\ref{eq.SC_PI_sol_e}) can be rewritten as:
\begin{equation}\label{eq.SC_PI_sol_e_+}
\dot{\rho}/q=\rho(1-\rho)(c/\Phi+\epsilon)(\rho/\rho_+-1)
\end{equation}
with $\rho_+=\rho_c(1+\Phi\epsilon/c)>\rho_s$. 
This value $\rho_+$ leads to an unstable equilibrium; In particular, $\dot{\rho}<0$ for $\rho<\rho_+$ so that $\rho\rightarrow\rho_c$ and hence eq.\ (\ref{eq.SC_PI_sol_e}) holds. 
\item
Finally when $\rho\ll\rho_c$ then $Q[\rho>\rho_c]\simeq0$ and eq.\ (\ref{eq.SC_PI_sol_e}) can be rewritten as:
\begin{equation}\label{eq.SC_PI_sol_e_-}
\dot{\rho}/q=\rho(1-\rho)(c/\Phi-\epsilon)(\rho/\rho_--1)
\end{equation}
with $\rho_-=\rho_c(1-\Phi\epsilon/c)<\rho_s$. As before, 
$\rho_-$ gives  an unstable equilibrium, because $\dot{\rho}<0$ for $\rho>\rho_-$ so that again $\rho\rightarrow\rho_c$ where eq.\ (\ref{eq.SC_PI_sol_e}) holds. 
\end{itemize}

\medskip

{\bf Remark 8:} 
In summary, the region $\rho_-<\rho<\rho_+$ becomes a finite basin of attraction for the dynamics. 
Note that when $\epsilon>c/(\alpha_c k_{max})$ then $\rho_+=\rho(0)$ has no solution 
and $\rho_c$ becomes the attractor in the whole $\alpha$ space. Our analysis thus shows that, for a range of initial densities of cooperators, 
there is a dynamical equilibrium characterized by an intermediate value of $\rho$, which is neither full defection nor full cooperation. 
Instead, for small enough or large enough values of $\rho(0)$ the system evolves towards the fully defective or fully cooperative Nash equilibrium, respectively.

\medskip

{\bf Remark 9:} 
The intuition behind the result above could be that mistakes can take a number of people away from the equilibrium, be it full defection or full cooperation, 
and that this takes place in a range of initial conditions that grows with the likelihood of mistakes.

\subsection{Best response}

Considering now the case of BR dynamics, the case of the coordination game is no different from that of the best-shot game and we cannot find rigorous proofs for our results, 
although we believe that we can substantiate them on firm grounds. To proceed, for this case eq.\ (\ref{eq.SS_BR}) becomes: 
\begin{equation}\label{eq.SC_BR}
\dot{\rho}/q=-\rho+Q[\pi_C>0]
\end{equation}
where we have taken into account that $\pi_D=0$ and $Q[\pi_C<\pi_D]=1-Q[\pi_C>\pi_D]$. 
Assuming that every node has degree $\bar{k}$, \emph{i.e.}, a regular random network, it is clear that there must be at least $[c/\alpha]+1$ neighboring cooperators in order to have $\pi_C>\pi_D$. 
Thus 
\begin{equation}Q[\pi_C>\pi_D]=Q[\pi_C>0]=\sum_{l=[c/\alpha]+1}^{\bar{k}}\binom{\bar{k}}{l}\rho^l(1-\rho)^{\bar{k}-l}\end{equation}
 and 
\begin{equation}\label{eq.SC_BR_1}
\dot{\rho}/q=-\rho+\sum_{l=[c/\alpha]+1}^{\bar{k}}\binom{\bar{k}}{l}\rho^l(1-\rho)^{\bar{k}-l}
\end{equation}
Once again, the difficulty is to show that $\rho_c=\rho(0)(\alpha_c/\alpha)$ is the unstable equilibrium. 
However, we can follow the same approach used with PI and write $Q[\pi_C>0]\simeq Q[\alpha\bar{k}\rho>c]=Q[\rho>\rho_c]$, \emph{i.e.}, we approximate $Q[\pi_C>0]$ as a Heaviside step function 
with threshold in $\rho_c$. We then again have three different cases as follows:
\begin{itemize}
\item
If $\rho\simeq\rho_c$ then $Q[\rho>\rho_c]\simeq 1/2$: We have $\dot{\rho}/q=-\rho+1/2$ and the attractor becomes $\rho\equiv1/2$. 
\item
If $\rho\gg\rho_c$ then $Q[\rho>\rho_c]\simeq 1$: We have $\dot{\rho}/q=-\rho+1$ and a stable equilibrium at $\rho\equiv1$. 
\item
Finally if $\rho\ll\rho_c$ then $Q[\rho>\rho_c]\simeq 0$: We have $\dot{\rho}/q=-\rho$ and a stable equilibrium at $\rho\equiv0$. 
\end{itemize}

\medskip

{\bf Remark 10:} 
As we may see, for  BR even without mistakes equilibria with intermediate values of the density of cooperators obtain in a range of initial densities. 
Compared to the situation with PI, in which we only found the absorbing states as equilibria, this points to the fact that more rational players would eventually converge to equilibria with higher payoffs. 
It is interesting to note that such equilibria could be related to those found by Galeotti {\em et al.} \cite{galeotti:2010} in the sense that not everybody in the network chooses the same action; 
However, we cannot make a more specific connection as we cannot detect which players choose which action---see, however, section \ref{sec.HMF_C_BR} below.

\bigskip

A similar approach allows some insight on the situation $\epsilon>0$. We start again from eq.\ (\ref{eq.SS_BR_e}), which now reduces to: 
\begin{equation}\label{eq.SC_BR_e}
\dot{\rho}/q=-(\rho-\epsilon)+Q[\pi_C>0](1-2\epsilon)
\end{equation}
Approximating as before $Q[\pi_C>0]\simeq Q[\rho>\rho_c]$ we again have the same three different cases. 
\begin{itemize}
\item 
If $\rho\simeq\rho_c$ then the attractor $\rho\equiv1/2$ is unaffected by the particular value of $\epsilon$. 
\item
If $\rho\gg\rho_c$ then the stable equilibrium lies at $\rho\equiv1-\epsilon$; 
\item
If $\rho\ll\rho_c$ then the stable equilibrium is at $\rho\equiv\epsilon$; 
\end{itemize}

\medskip

{\bf Remark 11:} 
Adding mistakes to BR does not change dramatically the results, as it did occur with PI. The only relevant change is that equilibria for low or high densities of cooperators are never homogeneous, 
as there is a percentage of the population that chooses the wrong action. Other than that, in this case the situation is basically the same 
with a range of densities converging to an intermediate amount of cooperators.

\section{Analysis of global welfare}\label{sec:welfare}

Having found the equilibria selected by different evolutionary dynamics, it is interesting to inspect their corresponding welfares (measured in term of average payoffs). 
We can again resort to the mean field approximation to approach this problem. 

\medskip

{\bf Best-shot game.} 
In this case the payoff of player $i$ is given by eq.\ (\ref{eq.SS}): $\pi_i=\Theta_H(y_i-1)-c\,x_i$. 
Within the mean field approximation, for a generic player $i$ with degree $k_i$ we can approximate the theta function as 
$\Theta_H(y_i-1)\simeq\rho+(1-\rho)[1-(1-\rho)^{k_i}]$, where the first term is the contribution given by player $i$ cooperating ($x_i=1$), 
whereas, the second term is the contribution of player $i$ defecting ($x_i=0$) 
and at least one of $i$'s neighbors cooperating ($x_j=1$ for at least one $j\in N_i$). It follows easily that:
\begin{equation}\label{eq.SSpi}
\langle\pi\rangle=\sum_k\,P(k)\,\{\rho+(1-\rho)[1-(1-\rho)^k]-c\rho\}.
\end{equation}
If $P(k)=\delta(k-\bar{k})$ (where $\delta(\cdot)$ stands for the Dirac delta function) then $\langle\pi\rangle=1-c\rho-(1-\rho)^{\bar{k}+1}$, 
whereas, if $P(k)=\bar{k}^k\,e^{-\bar{k}}/k!$ then $\langle\pi\rangle=1-c\rho-(1-\rho)e^{-\rho\bar{k}}$. 
We recall that in the simple case where players do not make mistakes ($\epsilon=0$), PI leads to a stationary cooperation level 
$\rho\equiv 0$, which corresponds to $\langle\pi\rangle=0$. On the other hand, with BR the stationary value of $\rho_s$ is given 
by eqn.\ (\ref{eq.SS_BR_1}) or (\ref{eq.SS_BR_2}), both leading to $\langle\pi_s\rangle=1-c\rho_s-\rho_s(1-\rho_s)$. 
As long as $\rho_s<c$, it is $\langle\pi_s\rangle>1-c$ (the payoff of full cooperation). 
We thus see that under BR players are indeed able to self-organize into states with high values of welfare in a non-trivial manner: 
defectors are not too many and are placed on the network such to allow any of them to be connected to at least one cooperator (and thus to get the payoff $\pi=1$); 
this, together with cooperators having $\pi=1-c$ by construction, results in a state of higher welfare than full cooperation.

\medskip

{\bf Coordination game.} 
Now player $i$'s payoff is given by eq.\ (\ref{eq:SC}): $\pi_i=(\alpha x_{N_i}-c)\,x_i$. 
Again within the mean field framework we approximate the term $x_{N_i}$ as $\rho k_i$, and we immediately obtain:
\begin{equation}\label{eq.SCpi}
\langle\pi\rangle=\sum_k\,P(k)\,\rho\{\alpha\rho k-c\}=\rho(\alpha\rho\bar{k}-c).
\end{equation}
$\langle\pi\rangle$ is thus a convex function of $\rho$, which (considering that $0\le\rho\le1$) attains its maximum value at $\rho=0$ 
when $\alpha<\alpha_{\pi}:=c/\bar{k}$, and at $\rho=1$ for $\alpha>\alpha_{\pi}$. Recalling that, in the simple case $\epsilon=0$, both with PI and BR 
there are two different stationary regimes ($\rho\rightarrow0$ for $\alpha \ll \alpha_c=c/[\rho(0)\bar{k}]$ and $\rho\rightarrow1$ for $\alpha \gg \alpha_c$), 
we immediately see that for $\alpha > \alpha_c > \alpha_{\pi}$ the stationary state $\rho=1$ maximizes welfare, and the same happens for $\alpha < \alpha_{\pi}$ 
with $\rho=0$. However, in the intermediate region $\alpha_{\pi}<\alpha<\alpha_c$ the stationary state is $\rho=0$ but payoffs are not optimal.

\section{Extension: Higher heterogeneity of the network}\label{sec:hmf}

In the two previous sections we have confined ourselves to the case in which the only information about the network we use is the mean degree, \emph{i.e.}, 
how many neighbors players do interact with on average. However, in many cases we may consider information on details of the network, such as the degree distribution, 
and this is relevant as most networks of a given nature (e.g., social) are usually more complex and heterogeneous than Erd\"{o}s-R\'{e}nyi random graphs. 
The heterogeneous mean field (HMF)~\cite{PastorSatorras2001} technique is a very common theoretical tool~\cite{Dorogovtsev2008} to deal with the intrinsic heterogeneity of networks. 
It is the natural generalization of the usual mean field (homogeneous mixing) approach to networks characterized by a broad distribution of the connectivity. 
The fundamental assumption underlying HMF is that the dynamical state of a vertex depends only on its degree $k$. In other words, all vertices having the same number of connections have exactly 
the same dynamical properties. HMF theory can be interpreted also as assuming that the dynamical process takes place on an annealed network~\cite{Dorogovtsev2008}, 
\emph{i.e.}, a network where connections are completely reshuffled at each time step, with the sole constraints that both the degree distribution $P(k)$ and the conditional probability $P(k|k')$ 
(\emph{i.e.}, the probability that a node of degree $k'$ has a neighbor of degree $k$, thus encoding topological correlations) remain constant. 

Note that in the following HMF calculations we always assume that our network is uncorrelated, \emph{i.e.}, $P(k'|k)=k'P(k')/\bar{k}$. 
This is consistent with our minimal informational setting, meaning that it represents the most natural assumption we can make.

\subsection{Best-shot game}

\subsubsection{Proportional imitation}

In this framework, considering more complex network topologies does not change the results we found before, and we again find a final state that is not a Nash equilibrium, namely full defection.

\medskip

{\bf Proposition 4.} 
In the HMF setting, under PI dynamics, when $\epsilon=0$ the final state for the population is the absorbing state with a density of cooperators $\rho=0$ (full defection) 
except if the initial state is full cooperation. 

\medskip

{\em Proof.} 
The HMF technique proceeds by building the $k$-block variables: We denote by $\rho_k$ the density of cooperators among players of degree $k$. The differential equation for the density of cooperators $\rho_k$ is: 
\begin{equation}\label{eq.SS_PI_h}
\dot{\rho}_k/q=(1-\rho_k)\sum_{k'}\rho_{k'}P(k'|k)\mathcal{P}_{D\rightarrow C}^{k\,k'}-\rho_k\sum_{k'}(1-\rho_{k'})P(k'|k)\mathcal{P}_{C\rightarrow D}^{k\,k'}
\end{equation}
The first term is the probability of picking a defector of degree $k$ with a neighboring cooperator of degree $k'$ times the probability of imitation 
(all summed over $k'$), whereas, the second term is the probability of picking a cooperator of degree $k$ with a neighboring defector of degree $k'$ 
times the probability of imitation (again, all summed over $k'$). For the best shot game, when $\epsilon=0$, we have:
$$\mathcal{P}_{C\rightarrow D}^{k\,k'}=c\qquad \mathcal{P}_{D\rightarrow C}^{k\,k'}=0\qquad\forall k,k'$$
We now introduce these values in eq.\ (\ref{eq.SS_PI_h}) and, using the uncorrelated network assumption, we arrive at:
\begin{equation}\label{eq.SS_PI_h_2}
\dot{\rho}_k/q=-c\rho_k\sum_{k'}(1-\rho_{k'})\frac{k'P(k')}{\bar{k}}=-c(1-\Theta)\rho_k
\end{equation}
where we have introduced the probability to find a cooperator following a randomly chosen link:
\begin{equation}\label{eq.SS_PI_h_Theta}
\Theta:=\sum_{k'}k'P(k')\rho_{k'}/\bar{k}.
\end{equation}
The corresponding differential equation for $\Theta$ reads
\begin{equation}\label{eq.SS_PI_h_3}
\dot{\Theta}=\sum_{k}kP(k)\dot{\rho}_k/\bar{k}=-qc\Theta(1-\Theta),
\end{equation}
and its solution has the same form of eq.\ (\ref{eq.SS_PI_sol}):
\begin{equation}\label{eq.SS_PI_h_sol}
\Theta(t)=\{1+[\Theta(0)^{-1}-1]e^{cqt}\}^{-1}
\end{equation}
with $\Theta(0)\equiv\rho(0)$ as $\rho_k(0)=\rho(0)$ $\forall k$. 
Hence $\Theta(t)\rightarrow0$ for $t\rightarrow\infty$ which implies $\rho_k(t)\rightarrow0$ for $t\rightarrow\infty$ and $\forall k$. $\blacksquare$

\medskip

{\bf Remark 12:} For the best-shot game with PI, the particular form of the degree distribution does not change anything. The outcome of evolution still is full defection, 
thus indicating that the failure to find a Nash equilibrium arises from the (bounded rational) dynamics and not from the underlying population structure. 
Again, this suggests that imitation is not a good procedure for the players to decide in this kind of games.

\bigskip

{\bf Proposition 5.} 
In the HMF setting, under PI dynamics, when $\epsilon\in(0,1)$ the final state for the population is the absorbing state $\rho=0$ (full defection) 
when $\epsilon < c$, $\rho=\rho(0)$ when $\epsilon=c$, and $\rho=1$ when $\epsilon>c$. When the initial state is $\rho(0)=0$ or $\rho(0)=1$, it remains unchanged.

\medskip

{\em Proof.} 
Eq.\ (\ref{eq.SS_PI_h}) is still valid, but now $\mathcal{P}_{C\rightarrow D}^{k\,k'}=c$ and $\mathcal{P}_{D\rightarrow C}^{k\,k'}=\epsilon$ $\forall k,k'$.
Again, using the uncorrelated network assumption, and introducing the effective cost $\tilde{c}=c-\epsilon$ we arrive at:
\begin{equation}\label{eq.SS_PI_h_2_2}
\dot{\rho}_k/q=-c(1-\Theta)\rho_k+\epsilon\Theta(1-\rho_k),
\end{equation}
\begin{equation}\label{eq.SS_PI_h_3_2}
\dot{\Theta}=-q\tilde{c}\,\Theta(1-\Theta),
\end{equation}
and at the end to a solution of the same form of eq.\ (\ref{eq.SS_PI_h_sol}):
\begin{equation}\label{eq.SS_PI_h_sol_2}
\Theta(t)=\{1+[\Theta(0)^{-1}-1]e^{\tilde{c}qt}\}^{-1}
\end{equation}
with $\Theta(0)\equiv\rho(0)$. 
Hence $\Theta(t)\rightarrow0$ for $t\rightarrow\infty$ (which implies $\rho_k(t)\rightarrow0$) only for $\tilde{c}>0$ ($\epsilon<c$); 
Instead for $\tilde{c}=0$ ($\epsilon=c$) then $\Theta(t)\equiv\Theta(0)$ ($\rho(t)\equiv\rho(0)$) $\forall t$, 
and for $\tilde{c}<0$ ($\epsilon>c$) then $\Theta(t)\rightarrow1$ for $t\rightarrow\infty$ (which implies $\rho_k(t)\rightarrow1$).
$\blacksquare$

\subsubsection{Best response}

Always within the deterministic scenario with $\epsilon=0$, for the case of best response dynamics the
differential equation for each of the $k$-block variables $\rho_k$ has the same form as eq.\ (\ref{eq.SS_BR}) above, 
where now to evaluate $Q_k[\pi_C>\pi_D]$ we have to consider the particular values of neighbors' degrees. 
As before, we consider the uncorrelated network case and introduce the variable $\Theta$ from eq.\ (\ref{eq.SS_PI_h_Theta}). We thus have
\begin{equation}\label{eq.SS_BR_ph}
Q_k[\pi_C>\pi_D]=\left[\sum_{k'}(1-\rho_k')P(k'|k)\right]^k=(1-\Theta)^k,
\end{equation}
and
\begin{equation}\label{eq.SS_BR_h}
\dot{\rho}_k/q=-\rho_k\,Q_k[\pi_C<\pi_D]+(1-\rho_k)\,Q_k[\pi_C>\pi_D]=(1-\Theta)^k-\rho_k.
\end{equation}
The differential equation for $\Theta$ is thus:
\begin{equation}\label{eq.SS_BR_h_3}
\dot{\Theta}/q=-\Theta+\sum_k(1-\Theta)^kkP(k)/\bar{k}
\end{equation}
whose solution depends on the form of degree distribution $P(k)$. Nevertheless, the critical value $\Theta_s$ such that 
the right-hand side of eq.\ (\ref{eq.SS_BR_h_3}) equals zero is also in this case the attractor of the dynamics.

\medskip

\begin{figure}[t!]
\centering
\includegraphics[width=0.5\textwidth]{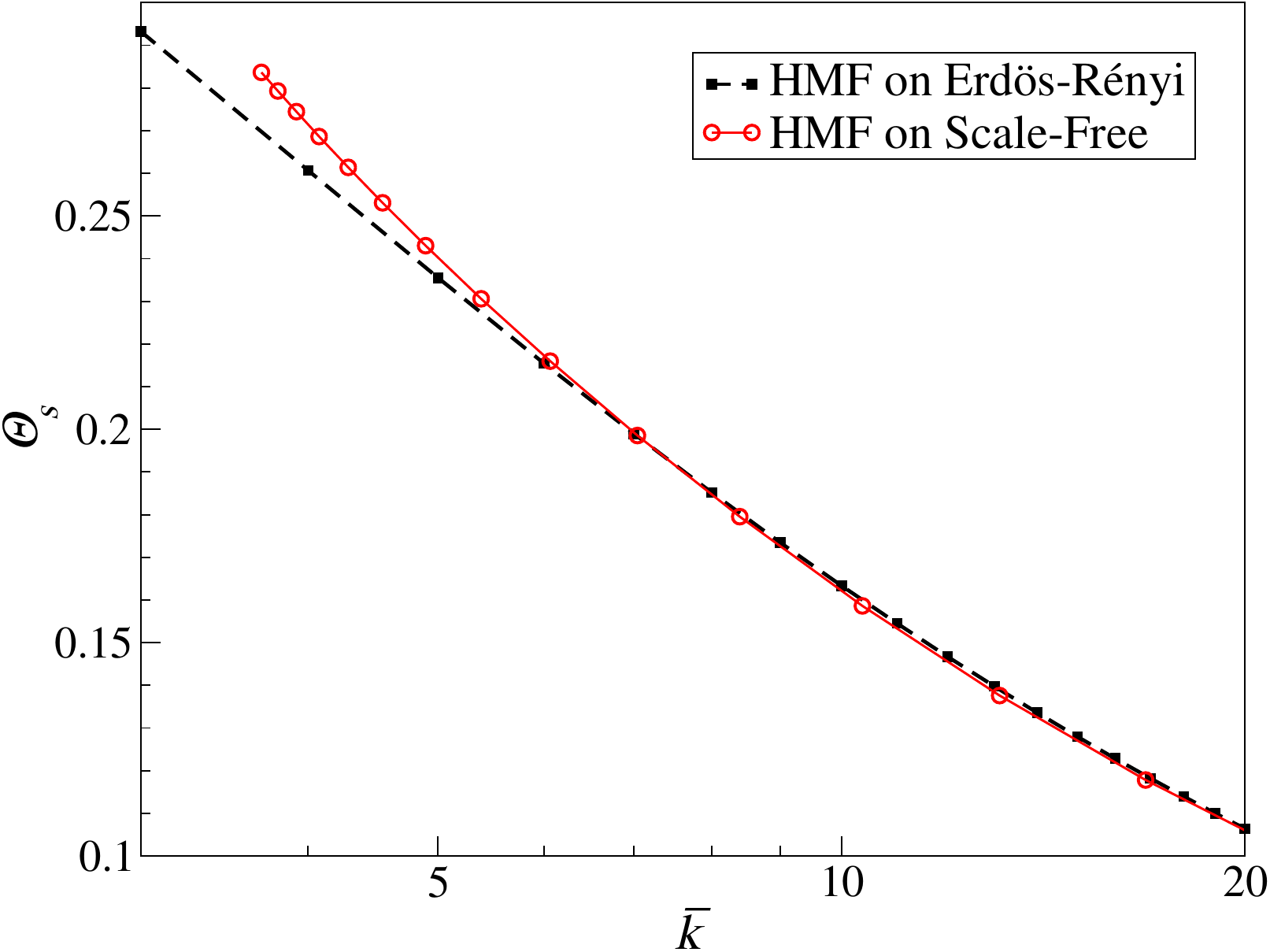}
\caption{Asymptotic value of the cooperator density for the best-shot game with BR dynamics for Erd\"{o}s-R\'{e}nyi and scale-free random graphs {(with $k_{min}=3$ and varying $\gamma$)}: 
$\Theta_s$ (main panel) and $\rho_s$ (inset) vs the average degree $\bar{k}$, obtained by numerically solving eq.\ (\ref{eq.SS_BR_h_3})}\label{fig.SS_BR_hmf}
\end{figure}

{\bf Remark 13:}
In order to assess the effect of degree heterogeneity, we have plotted in Fig.\ \ref{fig.SS_BR_hmf} the numerical solution for two random graphs, 
an Erd\"os-R\'enyi graph with a homogeneous degree distribution, and a scale-free graph with a much more heterogeneous distribution $P(k)=(\gamma-1)k_{min}^{(\gamma-1)}/k^\gamma$. 
In both cases, the networks are uncorrelated so our framework applies. As we can see from the plot, the results are not very different, and they become more similar as the average degree increases. 
This is related on one hand to the particular form of Nash equilibria for strategic substitutes, where cooperators are generally the nodes with low degree, 
and on the other hand to the fact that the main difference between a homogeneous and a scale-free $P(k)$ lies in the tail of the distribution. 
In this sense, the nodes with the highest degrees (that can make a difference) do not contribute to $\Theta_s$ and thus their effects on the system is negligible.

\bigskip

If we allow for the possibility of mistakes, the starting point of the analysis is---for each of the $k$-block variables $\rho_k$---the differential equation given by eq.\ (\ref{eq.SS_BR}). 
Recalling that $Q_k[\pi_C>\pi_D]=(1-\Theta)^k$, we easily arrive at:
\begin{equation}\label{eq.SS_BR_h_2}
\dot{\rho}_k/q=\epsilon-\rho_k+(1-2\epsilon)(1-\Theta)^k
\end{equation}
\begin{equation}\label{eq.SS_BR_h_3_2}
\dot{\Theta}/q=\epsilon-\Theta+(1-2\epsilon)\sum_k(1-\Theta)^kkP(k)/\bar{k}.
\end{equation}
A sufficient condition for the existence of a dynamical attractor $\Theta_s$ is $\epsilon<1/2$: also in heterogeneous networks, all reasonable values for the probability of errors 
allow for the existence of stable equilibria.

\subsection{Coordination game}

Unfortunately, for the coordination game, working in the HMF framework is much more complicated, and we have been able to gain only qualitative but important insights on the system's features. 
For the sake of clarity, we illustrate only the deterministic case in which no mistakes are made ($\epsilon=0$).

\subsubsection{Proportional imitation}

The average payoffs of cooperating and defecting for players with degree $k$ are:
$$\pi_D^k=0\quad\forall k\quad;\qquad\pi_C^k=\alpha k\left[\sum_{k'}P(k'|k)\rho_{k'}\right]-c=\alpha k\Theta-c$$
where $\Theta$ is the same as defined in eq.\ (\ref{eq.SS_PI_h_Theta}). 

We then use our starting point for the HMF formalism, eq.\ (\ref{eq.SS_PI_h}), where now the probabilities of imitation are:
$$\mathcal{P}_{D\rightarrow C}^{k\,k'}=(\pi_C^{k'}-\pi_D^k)Q[\pi_C^{k'}>\pi_D^k]/\Phi=(\alpha k'\Theta-c)\,Q_{k'}[\pi_C>0]/\Phi$$
$$\mathcal{P}_{C\rightarrow D}^{k\,k'}=(\pi_D^{k'}-\pi_C^k)Q[\pi_D^{k'}>\pi_C^k]/\Phi=-(\alpha k\Theta-c)\left\{1-Q_k[\pi_C>0]\right\}/\Phi$$
Once again within the assumption of an uncorrelated network, we find:
\begin{equation}\label{eq.SC_PI_h_2}
\Phi\dot{\rho}_k/q=(1-\rho_k)\sum_{k'}\rho_{k'}\frac{k'P(k')}{\bar{k}}(\alpha k'\Theta-c)Q_{k'}[\pi_C>0]+\rho_k\sum_{k'}(1-\rho_{k'})\frac{k'P(k')}{\bar{k}}(\alpha k\Theta-c)\left\{1-Q_k(\pi_C>0)\right\}
\end{equation}
In the second term we can carry out the sum over $k'$, which yields $\sum_{k'}(1-\rho_{k'})k'P(k')/\bar{k}=1-\Theta$. 
We are now ready to write the differential equation for $\Theta$:
\begin{eqnarray}\label{eq.SC_PI_h_3}
\Phi\bar{k}\dot{\Theta}/q=\sum_{k}kP(k)\dot{\rho}_k/q&=&\sum_{k}kP(k)(1-\rho_k)\sum_{k'}\rho_{k'}\frac{k'P(k')}{\bar{k}}(\alpha k'\Theta-c)Q_{k'}[\pi_C>0]\nonumber\\
&+&\sum_{k}kP(k)\rho_k(1-\Theta)(\alpha k\Theta-c)\left\{1-Q_k[\pi_C>0]\right\}.
\end{eqnarray}
Carrying out the summation over $k$ in the first term (which results again in $1-\Theta$), and relabeling $k'$ as $k$ we are left with
\begin{eqnarray}\label{eq.SC_PI_h_4}
\Phi\bar{k}\dot{\Theta}/q&=&\sum_{k}kP(k)\rho_{k}(1-\Theta)(\alpha k\Theta-c)Q_k[\pi_C>0]+\sum_{k}kP(k)\rho_k(1-\Theta)(\alpha k\Theta-c)\left\{1-Q_k[\pi_C>0]\right\}\nonumber\\
&=&\sum_{k}kP(k)\rho_{k}(1-\Theta)(\alpha k\Theta-c)=(1-\Theta)\Theta\left[\alpha\sum_{k}k^2P(k)\rho_{k}-c\bar{k}\right].
\end{eqnarray}
Finally,  introducing the new variable
\begin{equation}\label{eq.SS_PI_h_Theta2}
\Theta_2:=\sum_{k'}(k')^2P(k')\rho_{k'}/\bar{k}
\end{equation}
we arrive at:
\begin{equation}\label{eq.SC_PI_h_sol}
\Phi\dot{\Theta}/q=(1-\Theta)\Theta(\alpha\Theta_2-c)
\end{equation}

\medskip

\begin{figure}[t!]
\centering
\includegraphics[width=0.5\textwidth]{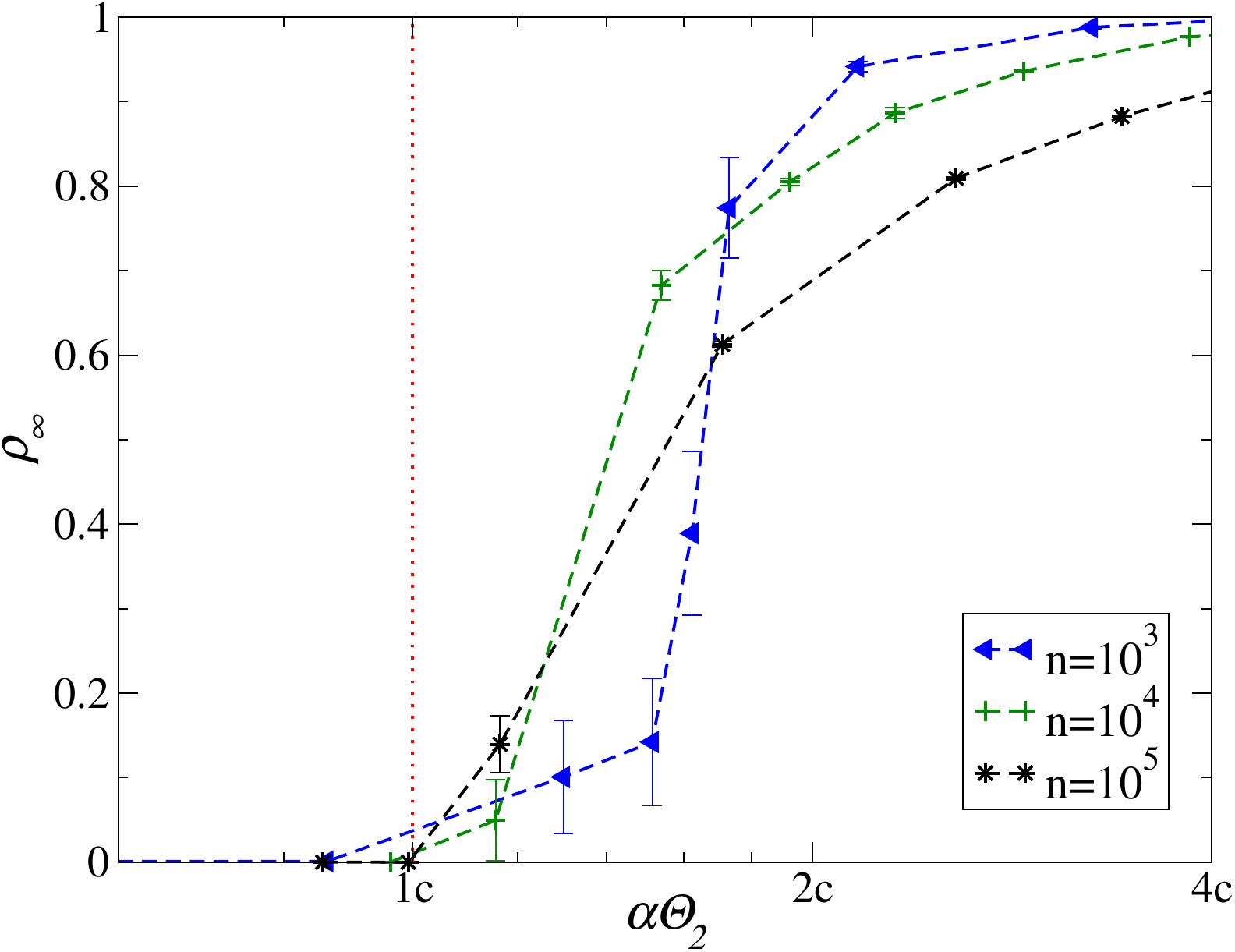}
\caption{Coordination game with PI dynamics for scale-free networks {(with $\gamma=2.5$, $k_{min}=3$ and $k_{max}=\sqrt{n}$)}: Stationary cooperation levels $\rho_\infty$ vs $\alpha\Theta_2$ 
for various system sizes $n$. The vertical solid line identifies the critical value of $\alpha_c\Theta_2=c$.}\label{fig.SC_PI_k2}
\end{figure}

{\bf Remark 14:}
While is it difficult to solve eq.\ (\ref{eq.SC_PI_h_sol}) in a self-consistent way, qualitative insights can be gained by defining $\alpha_c(t)=c/\Theta_2(t)$, 
and by rewriting eq.\ (\ref{eq.SC_PI_h_sol}) as $\dot{\Theta}(t)\simeq \alpha/\alpha_c(t)-1$ (the term $(1-\Theta)\Theta\ge0$ always and thus can be discarded). 
Now, starting from the beginning of the game at $t=0$, the initial conditions $\{\rho_{k}(0)\}$ univocally determine the value of $\Theta_2(0)$ and thus of $\alpha_c(0)$. 
For $\alpha<\alpha_c(0)$, $\dot{\Theta}(0)<0$ and $\rho$ decreases. Because of this, in the next time step $t=1$ we have on average that $\Theta_2(1)<\Theta_2(0)$, 
meaning $\alpha_c(1)>\alpha_c(0)>\alpha$: $\dot{\Theta}(1)<0$ again and $\rho$ keeps decreasing. By iterating such a reasoning, we conclude that in this case the stable equilibrium is $\Theta=0$. 
Symmetrically, for $\alpha>\alpha_c(0)$ the attractor becomes $\Theta=1$, and the transition between the two regimes lies at $\alpha\equiv\alpha_c(0)$.
Note that $\Theta_2$ is basically the second momentum of the degree distribution, where each degree $k$ is weighted with the density $\rho_k$. 
Recalling that $\bar{k^2}$ may diverge for highly heterogeneous networks (for instance, it diverges for scale-free networks with $\gamma<3$), 
and that for the coordination game cooperation is more favorable for players with many neighbors (hence $\rho_k\simeq1$ for high $k$), 
we immediately see that in these cases $\Theta_2$ diverges as well (as the divergence is given by nodes with high degree). 
Thus, while at the transition point the product $\alpha_c\Theta_2$ remains finite (and equal to $c$), $\alpha_c\rightarrow0$ to compensate for the divergence of $\Theta_2$ (Figure \ref{fig.SC_PI_k2}). 
We can conclude that, in networks with broad $P(k)$ and in the limit $n\rightarrow\infty$, cooperation emerges also when the incentive to cooperate ($\alpha$) vanishes. 
This is likely to be related to the fact that as the system size goes to infinity, so does the number of neighbors of the largest degree nodes. 
This drives hubs to cooperate, thus triggering a non-zero level of global cooperation.
However, if the network is homogeneous, neither $\bar{k^2}$ nor $\Theta_2$ diverge, so that $\alpha_c$ remains finite 
and the fully defective state appears also in the limit $n\rightarrow\infty$.

\subsubsection{Best response}\label{sec.HMF_C_BR}

For BR dynamics, we would have to begin again from the fact that the differential equation for each of the $k$-block variables $\rho_k$ has the same form of eq.\ (\ref{eq.SC_BR}). 
We would then need to evaluate $Q_k[\pi_C>0]=\sum_{l=[c/\alpha]+1}^k\binom{k}{l}[\sum_{k'}\rho_{k'}P(k'|k)]^l[1-\sum_{k'}\rho_{k'}P(k'|k)]^{k-l}=\sum_{l=[c/\alpha]+1}^k\binom{k}{l}\Theta^l(1-\Theta)^{k-l}$. 
As in the homogeneous case, such expression is difficult to treat analytically. 
Alternatively, we can perform the approximation of setting $Q_k[\pi_C>0]=Q[\alpha k\Theta>c]$, \emph{i.e.}, we approximate $Q_k[\pi_C>0]$ with a Heaviside step function 
with threshold in $\Theta=c/(\alpha k)$. This leads to:
\begin{eqnarray}
\dot{\rho}_k/q=-\rho_k & & \mbox{ for }k<c/(\alpha\Theta)\label{eq.SC_BR_h_1}\\
\dot{\rho}_k/q=-\rho_k+1& & \mbox{ for }k>c/(\alpha\Theta)\label{eq.SC_BR_h_2}
\end{eqnarray}
\begin{equation}\label{eq.SC_BR_h_3}
\dot{\Theta}/q=-\Theta+\sum_{k>c/(\alpha\Theta)}kP(k)/\bar{k}
\end{equation}
{and to the following self-consistent equation for the equilibrium $\Theta_s$:}
\begin{equation}\label{eq.SC_BR_h_4}
\Theta_s=\sum_{k>c/(\alpha\Theta_s)}kP(k)/\bar{k}
\end{equation}
whose solution strongly depends on the form of degree distribution $P(k)$. 
Indeed, if the network is highly heterogeneous (\emph{e.g.}, a scale-free network with $2<\gamma<3$), it can be shown that $\Theta_s$ is a stable equilibrium 
whose dependence on $\alpha$ is of the form $\Theta_s\sim\alpha^{(\gamma-2)/(3-\gamma)}$, \emph{i.e.}, there exists a non-vanishing cooperation level $\Theta_s$ no matter how small 
the value of $\alpha$. However, if the network is more homogeneous (\emph{e.g.}, $\gamma>3$), $\Theta_s$ becomes unstable and for $\alpha\rightarrow0$ the system always falls 
in the fully defective Nash equilibria. 
Another important characterization of such system comes from considering eq.\ (\ref{eq.SC_BR_h_1}) and eq.\ (\ref{eq.SC_BR_h_2}): 
We have $\rho_k(t)\rightarrow0$ when $k<c/(\alpha\Theta_s)$ and $\rho_k(t)\rightarrow1$ for $k>c/(\alpha\Theta_s)$. 
In this sense, we find a qualitative agreement between the features of our equilibria and those found by Galeotti {\em et al.} \cite{galeotti:2010}, 
in which players' actions show a monotonic, non-decreasing dependence on their degrees.

\section{Comparison with Numerical Simulations}\label{sec:simu}

Before discussing and summarizing our results, one question that arises naturally is whether, given that mean field approaches are approximations in so far as they assume interactions 
with a typical individual (or classes of typical individuals), our results are accurate descriptions of the real dynamics of the system. Therefore, in this section we present a brief comparison 
of the analytical results we obtained above with those arising from a complete program of numerical simulations of the system recently carried out by us, whose details can be found in \cite{us:2014} 
(along with many additional findings on issues that cannot be analytically studied). In this comparison, we focus on the scenario in which mistakes are not allowed ($\epsilon=0)$ as it, 
being deterministic, allows for a meaningful comparison of theory and simulations without extra effects arising perhaps from poor sampling. 

Concerning the best-shot game, numerical simulations fully confirm our analytical results. With PI, the dynamical evolution is in perfect agreement with that predicted by both MF and HMF 
theory---which indeed coincide when (as in our case) $\rho_k(t=0)$ does not depend on~$k$. Simulations and analytics agree well also when the dynamics is BR: The final state of the system is, 
for any initial condition, a Nash equilibrium with cooperators ratio $\rho_s$ (which decreases for increasing network connectivity). Yet, the $\rho_s$ solution of $\dot{\rho}=0$ from eq.\ (\ref{eq.SS_BR_2_e}) 
slightly underestimates the one found in simulations---probably because of the approximation made in computing the probabilities $Q$ of eq.~(\ref{eq.SS_BR}). 
Notwithstanding this minor quantitative disagreement, we can safely confirm the validity of our analytical results. 

On the other hand, the agreement between theory and simulations is also good for coordination games with PI dynamics. On homogeneous networks, numerical simulations show an abrupt transition 
from full defection to full cooperation as $\alpha$ crosses a critical value value $\alpha_T$.
The MF theory is thus able to qualitatively predict the behavior of the system; furthermore, while $\alpha_T$ is somewhat smaller than the $\alpha_c$ predicted by the theory, simulations also show
that $\alpha_T \rightarrow \alpha_c$ in the infinite network size, which implies that for reasonably large systems our analytical predictions are accurately fulfilled.
Finally, simulations cannot find other Nash equilibrium (with intermediate cooperation levels) than full defection, again as predicted by the MF calculations. 
On heterogeneous networks instead, simulations show a smooth crossover between full defection and full cooperation, and the point at which the transition starts ($\alpha_T$) tends to zero 
as the system size grows. Therefore, the most important prediction of HMF theory, namely that the fully defective state disappears in the large size limit (a phenomenon not captured by the simple MF approach), 
is fully confirmed by simulations. Finally, concerning BR dynamics for coordination games, we have a similar scenario: In homogeneous networks, simulations allow to find a sharp transition at $\alpha_T$ 
from full defection to full cooperation, featuring many non-trivial Nash equilibria (all characterized by intermediate cooperation levels) in the transient region. 
This behavior, together with $\alpha_T \rightarrow \alpha_c$ in the infinite network size, agrees well with the approximate theoretical results. Heterogeneous networks instead feature a continuous transition, 
and it appears from numerical simulations that---in the infinite network size---a Nash equilibrium with non-vanishing cooperation level exists no matter how small the value of $\alpha$, 
exactly as predicted by the HMF calculations. 

We can conclude that the set of analytical results we are presenting in this paper provides, in spite of its approximate character, a very good description of the evolutionary equilibria 
of our two prototypical games, particularly so when considering the more accurate HMF approach.

\medskip

\section{Conclusion}\label{sec:end}

In this paper, we have presented two evolutionary approaches to two paradigmatic games on networks, namely the best-shot game and the coordination game as representatives, respectively, 
of the wider classes of strategic substitutes and complements. As we have seen, using the MF approximation we have been able to prove a number of rigorous results, 
and otherwise to get good insights on the outcome of the evolution. 
Importantly, numerical simulations support all our conclusions and confirm the validity of our analytical approach to the problem.

Proceeding in order of increasing cognitive demand, we first summarize what we have learned about PI dynamics, equivalent to replicator dynamics in a well-mixed population. 
For the case of the best-shot game, this dynamics has proven unable to refine the set of Nash equilibria, as it always leads to outcomes that are not Nash. 
On the other hand, the asymptotic states obtained for the coordination game are Nash equilibria and constitute indeed a drastic refinement, selecting a restricted set of values for the average cooperation. 
We believe that the difference between these results arises from the fact that PI is an imitative dynamics and in a context such as the best-shot game, in which equilibria are not symmetric, 
this leads to players imitating others who are playing ``correctly'' in their own context but whose action is not appropriate for the imitator. 
In the coordination game, where the equilibria should be symmetric, this is not a problem and we find equilibria characterized by an homogeneous action.
Note that imitation is quite difficult to justify for rational players (as humans are supposed to act), because it assumes bounded rationality or lack of information 
leaving players no choice but copying others' strategies \cite{schlag:1998}. Indeed, imitation is much more apt to model contexts as biological evolution, where payoffs are interpreted 
as reproductive successes within natural selection \cite{MaynardSmith1995}. Under this interpretation, in the best-shot game for instance, it is clear that a cooperator surrounded by defectors 
would die out, and be replaced by the offspring of one of its neighboring defectors.

When going to a more demanding evolutionary rule, BR does lead by construction to Nash equilibria---when players are fully rational and do not make mistakes. 
We are then able to obtain predictions on the average level of cooperation for the best-shot game but still many possible equilibria are compatible with that value. 
Predictions are less specific for the coordination game, due to the fact that---in an intermediate range of initial conditions---different equilibria with finite densities of cooperators are found. 
The general picture remains the same in terms of finding full defection or full cooperation for low or high initial cooperation, but the intermediate region is much more difficult to study. 

Besides, we have probed into the issue of degree heterogeneity by considering more complex network topologies. Generally speaking, the results do not change much, at least qualitatively, 
for any of the dynamics applied to the best-shot game. The coordination game is more difficult to deal with in this context but we were able to show that, 
when the number of connections is very heterogeneous, cooperation may obtain even if the incentive for cooperation vanishes. 
This vanishing of the transition point is reminiscent of what occurs for other processes on scale-free networks, such as percolation of epidemic spreading~\cite{Dorogovtsev2008}. 
Interestingly, our results are in contrast with \cite{bram:2007}, in the sense that---for our dynamical approach---coordination games are more affected by the network 
(and are henceforth more difficult to tackle) that anti-coordination ones.

Finally, a comment is in order about the generality of our results. We believe that the insight on how PI dynamics drives the two types of games studied here should be applicable in general, 
\emph{i.e.}, PI should lead to dramatic reductions of the set of equilibria for strategic complements, but is likely to be off and produce spurious results for strategic substitutes, 
due to imitation of inappropriate choices of action. On the other hand, BR must produce Nash equilibria, as already stated, leading to significant refinements for strategic substitutes 
but to only moderate ones for strategic complements. This conclusion hints that different types of dynamics should be considered when refining the equilibria of the two types of games, 
and raises the question of whether a consistently better refinement could be found with only one dynamics. In addition, our findings also hint to the possible little relevance 
of the particular network considered on the ability of the dynamics to cut down the number of equilibria. In this respect, it is important to clarify that 
while our results should apply to a wide class of networks going from homogeneous to extremely heterogeneous, networks with correlations, clustering, or other nontrivial structural properties 
might behave differently. These are relevant questions for network games that we hope will attract more research in the near future.

\section*{Acknowledgments}

We are thankful to Antonio Cabrales, Claudio Castellano, Sanjeev Goyal, Angel S\'anchez and Fernando Vega-Redondo for their feedback on early versions of the manuscript 
and advice on the presentation of our results. 
This work was supported by the Swiss Natural Science Foundation (grant no PBFRP2\_145872) and the EU project CoeGSS (grant no 676547).

\section*{Competing interests}

The author declares that there is no conflict of interest regarding the publication of this paper

\end{document}